# Thermally-activated current transport in MgB$_2$ films

S. Patnaik*, A. Gurevich, S.D. Bu, J. Choi, C.B. Eom, and D.C. Larbalestier

Applied Superconductivity Center, University of Wisconsin, Madison, WI 53706

*School of Physical Sciences, JNU, New Delhi 110067, India

## Abstract

Thermally-activated flux flow (TAFF) resistivity above the irreversibility field B$_i$ is reported for two different c-axis textured MgB$_2$ superconducting films. Transport measurements at different perpendicular magnetic fields 0 < B < 9 T and temperatures from 5 to 40K reveal TAFF Ohmic resistivity ρ(T,B) for B > B$_i$ described by the Arrhenius law, ρ=ρ$_0$exp (-U/T) with the quadratic field dependence of the activation energy, U(B,T) = U$_0$(T)[1 - B/B$_{c2}$(T)]$^2$. Our transport measurements on bulk MgB$_2$ ceramic samples also show the TAFF behavior, but do not show the quadratic field dependence of U(T,B). We explain our results in terms of thermally-activated drift of pre-existing quenched dislocations in the vortex lattice. Our results indicate that thermal fluctuations can be essential in determining the irreversibility field in MgB$_2$ though to a much lesser extent than in high-temperature superconductors.

# I. Introduction

The "intermediate-$T_c$" superconductor MgB$_2$ [1] has brought to focus the physics of two gap superconductivity [2-4], as well as new possibilities for applications [5-7]. Supercurrent transport in MgB$_2$ in magnetic fields is characterized by the lack of apparent weak link behavior [8], and a comparatively moderate anisotropy of the upper critical field $B_{c2}^{\parallel}/B_{c2}^{\perp}$ ranging from 2 to 5, depending on the temperature T and specific material form [9-18] (the indices $\parallel$ and $\perp$ correspond to **B** parallel and perpendicular to the ab plane, respectively). These features of MgB$_2$ favorably distinguish it from the layered high-$T_c$ superconductors (HTS), for which the pronounced weak link behavior of grain boundaries, high anisotropy and giant thermally activated flux creep strongly limit the field region, B < $B_i$(T) in which HTS can carry supercurrents. Although MgB$_2$ also exhibits flux creep [19] and ohmic voltage-current (V-I) characteristics above the irreversibility field $B_i$(T), both $B_{c2}$ and $B_i$ can be significantly increased by alloying with nonmagnetic impurities and by irradiation [9, 20-27]. At the same time, MgB$_2$ exhibits electromagnetic behavior somewhat similar to that of HTS, such as broadening of the resistivity curve $\rho$(T, B) near $T_c$ [28-34] as the magnetic field [28-30] and current [31] are increased. In particular, a thermally activated flux flow (TAFF) resistivity was observed on textured bulk MgB$_2$ [30], and thermally-activated flux creep with the activation energy U proportional to the film thickness was reported for MgB$_2$ films [32].

These features of MgB$_2$ pose the fundamental question whether the difference between the magnetic field behavior of HTS and MgB$_2$ is merely quantitative, and whether the irreversibility field $B_i$(T) of MgB$_2$ may also have a thermally activated origin. The latter would imply the ohmic TAFF resistivity above $B_i$(T), just as is characteristic of HTS materials [35-38]. In this paper we address this issue by combined experimental and theoretical analysis of the temperature and field dependences of the ohmic TAFF resistivity $\rho$(T,B) of MgB$_2$ films found between $B_i(T)$ and $B_{c2}(T)$, which indeed shows the Arrhenius behavior, $\rho(T,B) = \rho_0 \exp[-U(T,B)/T]$ with a universal field dependence of the activation energy $U(T,B) = U_0(T)[1 - B/B_{c2}(T)]^2$. Thin films are particularly convenient to reveal thermally activated vortex dynamics, because the activation energy U decreases as the film thickness d is decreased below the pinning correlation length $L_c$ along the field direction [38-40]. The two-dimensional (2D) collective vortex dynamics and TAFF resistivity were indeed observed on low-$T_c$ weak-pinning Mo-Ge films and multilayers [41,42] and YBCO/PrBa$_2$Cu$_3$O$_7$ multilayers [43]. In this paper we report temperature and field dependencies of the TAFF resistivity on MgB$_2$ films.

The paper is organized as follows. First we present experimental data showing the ohmic voltage-current characteristic and TAFF resistivity of textured MgB$_2$ films which exhibit a clear Arrhenius behavior of $\rho(T,B) = \rho_0 \exp[-U(T,B)/T]$ with a quadratic field dependence, $U = U_0[1 - B/B_{c2}(T)]^2$, of the activation energy U above the irreversibility field $B_i \approx 0.8 B_{c2}$. Then we propose a model, which accounts for the observed $\rho(T,B)$ dependence by thermally activated hopping of quenched edge dislocations in the vortex lattice. The analysis is followed by a discussion on the effect of thermal activation on the irreversibility field $B_i(T)$ in MgB$_2$.

## II. Experimental details

Two strongly textured MgB$_2$ films were prepared using different processing techniques. Film 1 was prepared by pulsed laser deposition from a sintered MgB$_2$ target at room temperature onto (111) oriented single crystal SrTiO$_3$ substrates [9]. After deposition, Film 1 was annealed in a tantalum envelope inside an evacuated niobium tube at 950°C for 15 minutes. Magnesium pallets were included in the tube to prevent magnesium loss. X-ray diffraction exhibited a strong c-axis fiber texture with random in plane texture. The full width at half maximum of the 002 MgB$_2$ rocking curve was ~ 8° Film 2 was made epitaxially by depositing boron on (0001) Al$_2$O$_3$ by RF magnetron sputtering, followed by a post deposition anneal at 850°C in the presence of magnesium vapor [44]. X-ray diffraction and cross-sectional TEM revealed that the $MgB_2$ film was oriented with its *c*-axis normal to the (0001) $Al_2O_3$ substrate and with a 30° rotation in the *ab*-plane with respect to the substrate. Deposition was carried out at 5 mTorr argon at 500°C using a pure boron target. The thickness d of films 1 and 2 was about 0.5 μm and 0.4 μm, respectively.

Measurements of the film resistance as a function of temperature and magnetic field were carried out using a 9T Quantum Design PPMS. Resistance was measured in four-probe configuration with a dc current of 1 mA applied perpendicular to the magnetic field. Silver wires of 10 μm in diameter were used as leads for low resistance contacts placed unto the film by silver paste. For the temperature scans, magnetic field was held constant to the accuracy of 0.1mT and temperature was incremented in no-overshoot mode in small steps. Similarly, for the field scans, temperature was held constant and field was incremented at the rate of 10 mT/s in the steps of 100 mT. The onset of resistive superconducting transition temperature in zero field is found to be 37 K for Film 1 and 35.5 K for Film 2. The resistivity at 40K for film 1 and 2 is 38 and 6μΩcm, respectively.

### III. Results and Discussions

#### A. Experimental data

We analyze the temperature and the magnetic field dependence of the resistance R(T,B) in the range of magnetic fields $B_i(T) < B < B_{c2}(T)$, above the irreversibility field $B_i$, where the current-voltage characteristics are ohmic. Here the definition of $B_{c2}$ is shown in Fig. 1, and $B_i$ is defined as a field at which the critical current density $J_c(B)$ extrapolates to zero at the standard 1µV/cm electric field criterion. Our previous transport measurements gave $B_i \approx 0.8B_{c2}$ [9] for the film similar to Film 1. The results of our experiments are summarized in Figs. 1-7.

Fig. 1 shows the temperature dependence of R(T) for Film 1 at different magnetic fields applied perpendicular to the surface and thus to the ab plane. The field causes a noticeable shift of the onset of the resistive transition, as well as broadening of the R(T) curves as B increases. Though certainly not as pronounced, the qualitative behavior of R(T) is reminiscent of the dramatic broadening of the resistive transition in HTS. If this broadening in $MgB_2$ is of the same thermally activated origin as in HTS, there should be a transition from a highly nonlinear V-I curve with a finite critical current $I_c$ below the irreversibility field $B_i(T)$ to a weakly nonlinear V-I curve which becomes ohmic at low currents, V = RI and $B > B_i$ where the thermally-activated flux flow (TAFF) resistance is smaller than the flux flow resistance $R_F = R_n B/B_{c2}$ [36]. This transition manifests itself in the well-known curvature change in the logR-logI plot [38], which was indeed observed on our films. An example of this behavior is shown in Fig. 2 for Film 2 at B = 1T. As the temperature changes, a marked transition at the irreversibility temperature $T_i \approx$ 32.4K occurs, from a highly nonlinear V-I curve with a finite $I_c$ at $T < T_i(B)$ to a weakly nonlinear V(J) with a finite resistance above $T_i(B)$. As follows from Fig. 2, the reversible E-J curve for $T > T_i(B)$ exhibits two rather different ohmic resistivities $\rho_F$ and $\rho_T$ at high and low currents respectively, where $\rho_F = \rho_n B/B_{c2}$ is the flux flow resistivity, and $\rho_n$ is the normal state resistivity. The low-I resistivity $\rho_T$ is the TAFF resistivity, which is usually presented in the form, $\rho_T = \rho_F \exp[-U(T,B)/T]$, where U(T,B) is the activation energy determined by plastic deformation of correlated domains of the pinned vortex lattice in the pinning potential [38]. It is the properties of the TAFF resistivity, which will be addressed in this work.

To demonstrate the thermally-activated origin of $\rho_T(T,H)$, we present the Arrhenius plots of representative low-I resistance data in Fig. 3, which shows a good linear dependence of lnR(T) on 1/T over 3 decades in R. This behavior indicates the TAFF resistance, $R = R_n \exp[-$

U(T,B)/$k_B$T] in our MgB$_2$ films (hereafter we take the pre-factor as R$_n$ rather than R$_F$, which yields an inessential shift in U, as discussed below). The inset in figure 1 shows the zero field resistivity versus temperature up to 300 K for Film 2. The residual resistivity ratio for both the films is ~ 2.

To get further insight into the mechanisms of TAFF vortex dynamics in MgB$_2$, we focus on the field dependence of the activation energy U(T,B) extracted from the slopes of the curves in Fig. 4. A linear relationship between U$^{1/2}$ and applied field is evident. In Fig. 5 we verify the same analysis for Film 2 using a different procedure, where the inset shows field scans at constant temperature. Normalized resistance is then plotted as a function of $(1 - B/B_{c2})^2$ at T = 15 K. The results of this analysis yield a universal parabolic dependence, U(T,B)=U$_0$(T)[1 – B/B$_{c2}$(T)]$^2$ for both Film 1 and 2. This behavior was found to be fairly robust and independent of the way the barrier U is extracted. For example, Figs. 4 and 5 show U(B) obtained by two different methods: from the mean slope of lnR(T) for Film 1 (Fig. 4) and from U = $k_B$Tln(R$_n$/R) at T = 15K for Film 2 (Fig. 5). In both cases a good fit to U$^{1/2}$ ∝ 1 – B/B$_{c2}$ was observed. The magnitude of the activation energy U$_0$(0) can be evaluated by extrapolating the linear dependence in Fig. 4 up to the intersection with the vertical axis, which yields the value U$_0$ ≈ 2×10$^3$ K, consistent with the scale of activation energy extracted from flux creep measurements on a 0.4 µm MgB$_2$ film [32].

Since U$_0$ >> $k_B$T$_c$, the behavior of U(T,B) obtained from the above analysis is rather insensitive to the choice of the pre-factor in the TAFF resistance. Indeed, both the pre-factor R$_n$ and the activation energy U can be re-scaled to R$_n^/$ and U$^/$ = U - $k_B$Tln(R$_n$/R$_n^/$), respectively without changing the observed resistance R = R$_n$exp(-U/$k_B$T). In particular, if R$_n^/$ = R$_F$, then U$^/$ = U – $k_B$Tln(B$_{c2}$/B), so for U ~ 10$^3$K, the difference between U and U$^/$ is only few percent. We also found that the behavior of U(T,B) is sensitive to the sample geometry. Indeed, the parabolic dependence, U ∝ $(1 – B/B_{c2})^2$ was only observed on thin films, whereas the Arrhenius plot for bulk sintered MgB$_2$ samples shown in Fig. 6 revealed a curvature of lnR(T). Possible mechanisms behind the difference between thin film and bulk samples are discussed below.

### B. Thermal depinning of vortex lattice

Our experimental data indicate that thermal fluctuations of vortices do contribute to the resistive behavior of MgB$_2$ above B$_i$. The key parameters of these thermal fluctuations can be estimated from the mean-squared amplitude of thermal vortex displacements $u^2$(T,B) in a uniaxial superconductor for **B**||c [46]:

$$u^2 = k_B T \left( \frac{16\pi\mu_0 \lambda^4 \eta}{B_{c2}\phi_0^3 g(b)} \right)^{1/2}, \tag{1}$$

$$g(b) = h(1-h)^3 \ln[2 + (2h)^{-1/2}], \tag{2}$$

where $\lambda$ is the in-plane London penetration depth, $\eta = B_{c2}^{\parallel}/B_{c2}^{\perp} > 1$ is the anisotropy parameter, $\phi_0$ is the flux quantum, h = B/B$_{c2}$(T) is the reduced magnetic field, and k$_B$ is the Boltzman constant. Following the approach developed for HTS [37,45], we define the thermal depinning field B*(T) at which $u$(T,B*) equals the in-plane coherence length $\xi$(T). This condition ensures thermal smearing of the pinning potential for vortex core pinning by point defects. From $u^2$(T,B*) = $\xi^2$(T) and $\xi$(T) = $\xi_0$(1 – t$^2$)$^{-1/2}$, we find that the dependence of the depinning field h = B*(T)/B$_{c2}$(T) on temperature t = T/T$_c$ is determined by the following equation

$$t^2 = g(h)/[\alpha^2 + g(h)], \tag{3}$$

Here the parameter $\alpha$ quantifies the strength of vortex thermal fluctuations,

$$\alpha = \frac{4\sqrt{2}\mu_0\pi\eta\kappa^2\xi_0^2 k_B T_c}{\phi_0^2}, \tag{4}$$

and $\kappa = \lambda/\xi$ is the Ginzburg-Landau parameter. For characteristic of $MgB_2$ values, $T_c = 40K$, $\kappa = 30$, $\xi_0 = 5$ nm [4], and $\eta \approx 2\text{-}6$, Eq. (4) yields $\alpha$ between 0.03 and 0.09. The small values of $\alpha$ indicate that vortex fluctuations in $MgB_2$ are weaker than in HTS, although $\alpha$ in $MgB_2$ could be increased by impurities [9]. For $\alpha \ll 1$ and $T_c - T \gg \alpha^2 T_c$, Eqs. (1)-(2) give $B^*(t)/B_{c2}(t) \approx 1 - [\alpha^2 t^2/(1-t^2)]^{1/3}$.

The calculated depinning line $B^*(T)$ is shown in Fig. 7 for $\alpha = 0.05$, 0.1, and 0.3. The curve $B^*(T)$ is two-valued because thermal fluctuations are most pronounced near the critical fields $B_{c1}(T)$ and $B_{c2}(T)$, where the shear modulus $C_{66}(B)$ of the vortex lattice vanishes. The $B^*(T)$ curve does not extrapolate to $T_c$ but has an infinite slope at the temperature $T_m < T_c$, where $B^*(T_m) \approx B_{c2}(T_m)/4$, and

$$T_m = \frac{T_c}{\sqrt{1+\alpha_0^2}}, \qquad \alpha_0^2 \approx \frac{256\alpha^2}{27\ln(2+\sqrt{2})} \tag{5}$$

For $\alpha = 0.08$, Eqs. (4) and (5) give $T_m \approx 0.975 T_c$. The depinning field $B^*(T)$ exhibits the same quantitative behavior as the observed irreversibility line $B_i(T)$ previously measured in the range of $T_c$ down to about 20 K [9]. However, for a typical value of $\alpha = 0.05$, the calculated curve $B^*(T)$ lies considerably higher than the observed $B_i(T)$. This fact indicates that melting of the vortex solid is irrelevant to $B_i(T)$ in $MgB_2$, because the melting line $B_m(T)$ defined by the usual criterion $u^2(T,B_m) = c_L^2 a^2(B_m)$ lies even higher than depinning line calculated from $u^2(T,B^*) = \xi^2(T)$. Here $a = (\phi_0/B)^{1/2}$ is the intervortex spacing, and $c_L \approx 0.3$ is the Lindemann constant [38].

Therefore, $B_m(T)$ is rather close to $B_{c2}$, so the vortex solid (disordered vortex lattice) in MgB$_2$ exists practically in the entire reversible field range $B_i < B < B_{c2}$. Furthermore, the fact that MgB$_2$ films exhibit a linear ohmic resistivity at $B_i < B < B^*$ indicates that the Lorentz force of transport current causes local plastic flow of the pinned vortex solid, as discussed below.

### C. Thermally-activated creep of vortex dislocations.

The Arhenius temperature dependence of the resistivity can be understood in terms of the TAFF theory in which [35-38]

$$\rho = \rho_0 \exp[-U(T,B)/k_B T], \qquad (6)$$

where we choose $\rho_0 = \rho_n$, as discussed above. The activation energy $U(T,B)$ is determined by a characteristic energy of local plastic deformations of the vortex solid, which can be interpreted in terms of thermally activated glide of dislocations. Under the action of the Lorentz force, these dislocations hop between neighboring stable positions in the vortex lattice, resulting in the TAFF resistivity. The dependence of $U$ on $T$ and $B$ is determined by two principal mechanisms by which the dislocations are generated: 1. Thermally activated excitation of dislocation pairs, 2. Thermally activated creep of pre-existing quenched dislocations.

The thermally activated generation of dislocation pairs results in $U_p(T,B) \approx (da^2 C_{66}/\pi)\ln(B_{c2}/B)$ for a thin film of thickness $d$ [47,48]. Using the expression for the shear modulus, $C_{66} = (B\phi_0/64\pi^2\lambda^2)(1 - B/B_{c2})^2$ [46], we get the quadratic field dependence $U_p \propto (1 - B/B_{c2})^2$, in agreement with our experimental data. Thermal dissociation of dislocation pairs is essential in layered HTS, for which this model describes well the TAFF resistivity in Bi-Sr-Ca-Co-O single crystals at low fields $B < 0.2B_{c2}$ where vortex pancakes are decoupled, and $d$ can be regarded as a thickness of the double layer between Cu-O planes [49]. However, this mechanism is far less effective for the weakly anisotropic MgB$_2$ films with $\lambda_0$ = 140 nm [4], and d = 500 nm, for which it gives the activation energy scale $U_p(0,0) = (d\phi_0^2/64\pi^3\lambda_0^2 k_B)\ln(B_{c2}/B) \approx 36730$ K, more than an order of magnitude higher than the observed values $U_0(0) \sim 2000$ K in Fig. 4, and the activation energies extracted from flux creep experiments on MgB$_2$ films with d $\cong$ 0.4-0.5 µm [32]. Thus, thermal activation of dislocations pairs can be ruled out, as it would give $\rho(T)$ many orders of magnitude smaller than the observed TAFF resistivity.

Now we discuss the second mechanism of TAFF due to plastic flow of pre-existing dislocations. Quenched dislocations in the vortex lattice are very common defects, which have been observed in many local probe experiments on various superconductors, starting from the classic decoration experiments by Träuble and Essmann [50]. More recently vortex dislocations in HTS have been revealed by means of micro Hall probes and the electron Lorentz microscopy (see, e.g., [51-53] and references therein). Dislocations have also been studied theoretically [54-57] and observed in molecular dynamics simulations of pinned flux line lattice [58-60]. The high density of quenched dislocations can be either produced by plastic deformation of the vortex lattice by pinning potential [55], or as a result of the formation of metastable vortex structures during remagnetization under a finite magnetic ramp rate. In the latter case dislocations could appear due to partial penetration or exit of vortex rows near the sample surface to sustain a finite flux gradient $B^{/} = \mu_0 J_c(B)$ in the critical state.

Quenched dislocations manifest themselves in the TAFF resistivity because they give rise to plastic flow of the vortex lattice analogous to the usual plastic deformation of crystalline solids under stress [61]. For further qualitative analysis, we do not distinguish between a macroscopically uniform distribution of dislocations or a polycrystalline vortex structure where dislocations are mostly located in the network of grain boundaries. We first estimate U for a single edge dislocation in a film of thickness d smaller than the pinning correlation length $L_c$, so that vortices are straight and perpendicular to the film surface. In the case of sparsely distributed pinning centers, the Lorentz force causes local plastic deformation due to slippage of dislocations between the pins. This mechanism determines the minimum energy barrier $U_i(T,B)$ controlled by intrinsic pinning of the dislocation by the Peierls potential, for which

$$U_i \cong \frac{db^2}{2\pi} e^{-4\pi l/b} C_{66}(T,B) \propto (1 - B/B_{c2})^2 . \tag{7}$$

Here $b \sim a$ is the Burgers vector along the glide direction of the dislocation, and $l \approx b/2$ is the width of the dislocation core in the Peierls theory [61]. As follows from Eq. (7), the model of pre-existing dislocations also gives the quadratic field dependence of $U(B) = U_0(T)[1 - B/B_{c2}(T)]^2$ observed in our experiment. However, the activation barrier $U_i(0,0) \approx U_p(0,0)\exp(-2\pi)/2 \approx 34K$ for pre-existing dislocations is much smaller than $U_p$ for the dislocation pairs, and the observed TAFF activation energy U. The difference between $U_i$ and U indicates that not only a more detailed theory of the dislocation core structure may be necessary, but also that pinning of vortex dislocations by microstructural defects is much stronger than by the Peierls potential. Because pinning increases U, it would make $U_d$ for the dislocation pairs even higher, but it would increase $U_i$ for quenched dislocations toward the observed values of U. Pinning of vortex dislocations is determined by complicated collective interaction between dislocations and pins [57], so we just estimate U due to pinning of an edge dislocation by two strong defects spaced by $l_i$. In this case the transport current causes the dislocation to form an arc between the pins. The dislocation gets depinned by the Frank-Reed mechanism [61] if the radius of the arc becomes of order $l_i$. The energy barrier U is then equal to the dislocation line energy times the length difference between the curved and the straight dislocation, whence

$$U \cong \frac{b^2 C_{66} l_i}{2\pi} \ln \frac{l_i}{a}. \qquad (8)$$

For $\lambda_0 = 140$ nm, characteristic of MgB$_2$ we get the line dislocation energy $\varepsilon_d = b^2 C_{66}/2\pi = 36.7$ K/nm. Thus Eq. (8) yields that the observed activation energy $U_0 \sim 2\times10^3$ K would correspond to an average pin spacing $l_i \sim 50$-$60$ nm.

For further qualitative analysis of the temperature dependence of R(T,B), we assume that pinning can account for the difference between $U_i$ and the observed U, and take the temperature dependences of $\lambda(T)$ and $B_{c2}(T)$ in the form: $\lambda(t) = \lambda_0(1 - t^2)^{-1/2}$ and $B_{c2}(t) = B_{c2}(0)(1 - t^2)/(1 + 0.4t^2)$. These interpolation formulas for $\lambda(t)$ and $B_{c2}(t)$ provide the correct Ginzburg-Landau behavior near $T_c$ and account for leveling off $\lambda(t)$ and $B_{c2}(t)$ at low T. In particular, the formula for $B_{c2}(t)$ approximates the well-known de-Gennes-Maki dependence of $B_{c2}(t)$ in dirty one-gap superconductors to the accuracy better than 5% and provides the correct relation $B_{c2}(0) = 0.7 B_{c2}' T_c$ where $B_{c2}' = |\partial B_{c2}/\partial T|_{Tc}$. Then Eqs. (6)-(7) yield the following temperature and field dependence of $\rho(T,B)$:

$$\rho = \rho_0 \exp\left[ -\frac{U_0}{k_B T_c}\left(\frac{1}{t} - t\right)\left(1 - \frac{h_0(1+0.4t^2)}{1-t^2}\right)^2 \right]. \qquad (9)$$

Here $U_0$ is the activation energy at T = 0, and $h_0 = B/T_c B_{c2}'$. The pre-factor $\rho_0 \cong \rho_n n_d \phi_0/B_{c2}$ is analogous the Bardeen-Stephen resistivity in which the vortex density is replaced with the density of quenched dislocations $n_d$ [49]. The fit of Eq. (9) to the experimental data of R(T) (Film 1) shown in Fig. 7 indicates that the model describes well the observed temperature dependences of R(T,B) in our films.

The parameter $U_0(0)/T_c$ extracted from the fits in Fig. 7 equals 52, 48, and 43 for the fields 7, 8 and 9 T, respectively. This variation of $U_0(0)/T_c$ may be due to the fact $\rho(T,B)$ was measured in a rather wide ($0.2T_c < T < T_c$) temperature range, $\rho(T,B)$ curves shifting to lower temperatures as H increases. In this case details of the dependencies of $\lambda(T)$ and $B_{c2}(T)$ in a broad temperature range become essential, so the conventional interpolation formula $\lambda(t) = \lambda_0(1 - t^2)^{-1/2}$ may not provide the necessary accuracy. A more rigorous approach can be based on the following expression for the penetration depth in dirty two-gap superconductors [62,63]

$$\frac{1}{\lambda^2} = \left(\frac{2\pi e}{c}\right)^2 \left(N_1 D_1 \Delta_1 \tanh\frac{\Delta_1}{2T} + N_2 D_2 \Delta_2 \tanh\frac{\Delta_2}{2T}\right), \qquad (10)$$

where the indices 1 and 2 correspond to $\sigma$ and $\pi$ bands of $MgB_2$, $N_{1,2}$ are partial densities of states, $\Delta_{1,2}$ are the superconducting gaps, $D_1$ and $D_2$ are electron diffusivities in $\sigma$ and $\pi$ bands due to intraband impurity scattering, e is the electric charge, and c is the speed of light. As follows from Eq. (10), the temperature dependence $\lambda(T)$ is not entirely determined by that of $\Delta(T)$ as in usual one-gap superconductors, but also by the relative contribution of two bands, which can vary significantly depending on which band has stronger scattering. Another reason for variation of $U_0(0)/T_c$ may be due to known deviation of $B_{c2}(T)$ in two-gap superconductors from the conventional deGennes-Maki formula [9,63]. Because of these complicated manifestations of two-gap superconductivity in $MgB_2$, the comprehensive comparison of the dislocation model with experiment would require direct measurements of $\lambda(T)$ and $B_{c2}(T)$ on the same film. However, given the excellent agreement with the quadratic field dependence of U(B) in Figs. 4 and 5, and the good fit in Fig. 7 with only moderate change in $U_0(0)/T_c$, we can conclude that the dislocation model does capture the essential thermally-activated behavior of R(T,B) in our films.

The dislocation model also indicates that the simple Eq. (7) may not work in bulk samples, as seen in Fig. 6. Indeed, for bulk samples, vortex dislocations do not hop as a whole, but first form a double-kink segment of length L(T,B), which then gets into the neighboring valley of the Peierls potential and then propagates sideways [61]. For the 3D vortex lattice, the thickness d in Eq. (7) should therefore be replaced by the length L(T,B) which is determined by elastic moduli of the vortex lattice, and essentially depends on T and B. For example, measurements of the thickness dependence of the flux creep activation energy indicate that the crossover length L(T,B) could be as large as 1mm [32], in which case even the Peierls potential could provide the observed $U_0 \sim b^2 C_{66} L \exp(-2\pi)/2\pi \sim 7 \times 10^4$ K for L = 1mm and $\lambda_0$ = 140 nm. In this case pinning of screw dislocations, for which the Peierls potential is absent [54,56] can become the limiting TAFF mechanism in bulk samples. Multiple dislocation mechanisms of TAFF and the additional temperature and field dependencies due to L(T,B) make the behavior of U(T,B) in bulk samples much more complicated, let alone the fact that the interpretation of the resistive transition in our bulk $MgB_2$ ceramic materials, is also complicated by percolation effects due to inhomogeneities and misoriented anisotropic grains [64-66].

Percolative effects due to inhomogeneous flux flow resistivity cannot produce the ohmic TAFF behavior for $B > B_i$. Indeed, let us consider the flux flow resistivity $\rho(\mathbf{r}) = \rho_n B/B_{c2}(\mathbf{r})$ with local $B_{c2}(\mathbf{r}) = \langle B_{c2} \rangle + \delta B_{c2}(\mathbf{r})$, where $\langle B_{c2} \rangle$ is a mean value, and $\delta B_{c2}(\mathbf{r})$ is a randomly inhomogeneous correction due to variations local $T_c$, mean free path, etc. Solving for distributions of electric field and currents, we can obtain the global resistivity $\langle \rho \rangle = \rho_n B/B_0(T)$, where the scaling field $B_0(T)$ can be calculated using, for example, the effective medium theory [64]. The resulting $B_0(T)$ depends on statistical correlation properties of inhomogeneities, but is generally of the order of the global $\langle B_{c2} \rangle$ and has a similar (usually weaker temperature dependence). Thus, $\langle \rho \rangle$ always remains linear in B so the flux flow model cannot possibly explain the exponential field dependence of the ohmic resistance R(B) of our films above $H_{irr}$.

Nor can this model explain the temperature dependence of R(T): as follows from Fig. 2, the low-J ohmic resistance R(T) at a fixed B = 1T drops by more than 2 orders of magnitude as T decreases from 34 to 32.8K. These features of R(T,B) clearly indicate very strong exponential temperature and field dependencies characteristic of TAFF, but not much weaker dependencies produced by randomly-inhomogeneous flux flow. Inhomogeneities could indeed become important below the global $B_i$ where separated flux flow "islands" in which local $B_i(\mathbf{r})$ is smaller than B, can affect the V-J characteristics [64]. However, this regime corresponds to the nonlinear part of the V-J curve in Fig. 2, but not to the ohmic TAFF state addressed in our work.

The reversible transport behavior of $MgB_2$ films at $B > B_i \approx 0.8\text{-}0.9B_{c2}$ indicates that elecrodynamics of our $MgB_2$ films turns out to be intermediate between that of HTS and low-$T_c$ superconductors like $Nb_3Sn$ in which the irreversibility field $B_i(T) < B_{c2}(T)$ has also been observed [67]. There is also an alternative interpretation of the transition at $B = B_i(T)$ in $MgB_2$ single crystals as being due to surface superconductivity [31,68]. In any case, there is a significant difference between $B_i \approx 0.8\text{-}0.9B_{c2}$ in our films and the onset of irreversible behavior $B_i$, in good single crystals which is much closer to $B_{c2}$ [68-70]. We believe that more experiments are needed to unambiguously clarify the reasons behind the difference in $B_i$ in films and single crystals. Here we briefly discuss several mechanisms, which might account for this difference. First, thermal activation of vortices in films are much stronger than in bulk samples, because the TAFF activation energy U decreases linearly as the film thickness d decreases. Defining the irreversibility field by the condition $U(B_i,T) = CT$, where C is a numerical parameter of the order of unity [38], and using Eq. (7), we obtain

$$\mathbf{B_i(T)} = \left[1 - \left(\frac{CT}{U_0(T)}\right)^{1/2}\right]\mathbf{B_{c2}(T)} \qquad (11)$$

Because $U_0$ is proportional to the film thickness (see Eq. (7)), $B_i$ in thin films with $d < L_c(T,B)$ can be smaller than in bulk samples.

Another possible reason for different $B_i$ in films and single crystals could be due to the effect of the sample geometry. Magnetic geometrical barrier can significantly affect the onset of bulk irreversible behavior in high quality single crystals with weak pinning, as has been shown by changing the cross-sectional shape of and $MgB_2$ single crystals [68] and Bi-2212 single crystals [71]. However, we believe that the surface superconductivity, which occurs at the field $H_{c3} = 1.69H_{c2}$ parallel to the flat sample surface, has no effect on transport behavior in our high-$J_c$ thin films in a *perpendicular* field. The critical current density $J_c \sim 100$ KA/cm$^2$ in our films is two orders of magnitude higher that that for the $MgB_2$ single crystals of Refs. 68-70. Furthermore, the thickness of our films d ~0.5 μm is $10^2$ times smaller typical thickness of single crystals of Refs. 68-70, so bulk pinning in our films is much stronger than the hysteretic effect of geometrical barrier, and the high demagnetizing factor rules out the surface superconductivity. Because bulk pinning in our films certainly dominates over the geometrical barrier, we can define the irreversibility field $B_i$ in a usual was as a field at which $J_c(T,B)$ measured at 1 μV/cm electric field criterion extrapolates to zero. However, it would be difficult to implement this conventional procedure in good single crystals because $J_c$ is low, so the geometrical barrier can strongly mask weak bulk pinning, as has been shown in Ref. 71. By contrast, the characteristic transport transition from a nonlinear to the ohmic resistive state in our films shown in Fig. 2 clearly indicates the thermally-activated flux flow behavior.

## Acknowledgements

This work was supported by NSF under the MRSEC on nanostructured materials and interfaces (DMR 9614707).

**Figure captions**

Fig 1. Zero field and in-field temperature dependences of electrical resistance for Film 1. External magnetic field varies from 0 to 9T in steps of 1T and is applied perpendicular to the film plane. Inset shows zero field resistivity curve R(T) up to room temperatures for Film 2. The figure also shows how we have defined the upper critical field $B_{c2}$ (T) and irreversibility field $B^*$ (T)

Fig. 2. Resistance R = V/I as a function of current density for B = 1T and different temperatures for Film 2. The transition from the nonlinear V(J) curve to the ohmic V-J curve at low J occurs at the temperature $T^*(B) \approx 32.4K$ above which $\rho(J)$ approaches a constant TAFF resistivity $\rho_T$ as $J \rightarrow 0$.

Fig 3. Arrhenius plot of the electrical resistance of $MgB_2$ film1 for magnetic field 1 < B < 9T perpendicular to film plane. The activation energy is determined from the slope in the linear region, U = - dlnR/dT.

Fig 4. A parabolic dependence of the activation energy $U(B) \propto (1 - B/B_{c2})^2$. The straight line is a guide to the eye and the data points are the magnitude of the slopes at different fields taken from the linear region of the Arrhenius plots in Fig. 3.

Fig 5. Normalized resistance is plotted as a function of $(1 - B/B_{c2})^2$ at T = 15 K for Film 2 to verify the parabolic field dependence of U(B). The inset shows the resistance as a function of external field at 17 K and 15K.

Fig 6. Arrhenius plot of a bulk sample at representative fields of 9, 8, 6, 4, and 2 T.

Fig 7. Resistance as a function of temperature for three representative fields 7, 8, and 9T for Film 1. The solid lines are fitting with Eq. (9) with the normal state resistance $R_n$ = 0.19Ω, $B_{c2}^{/}$ = 0.45T/K [9], $T_c$ = 37 K, and $U_0(0)/T_c$ = 52, 48 and 43, respectively. The dotted line represents the irreversibility field cut off at 1μV/cm electric field criterion.

Fig 8. The depinning field $B^*(T)$ (solid lines) calculated from Eq. (3) for different values of α. The dashed curve shows $B_{c2}(T)$.

Fig.1 Patnaik, S.

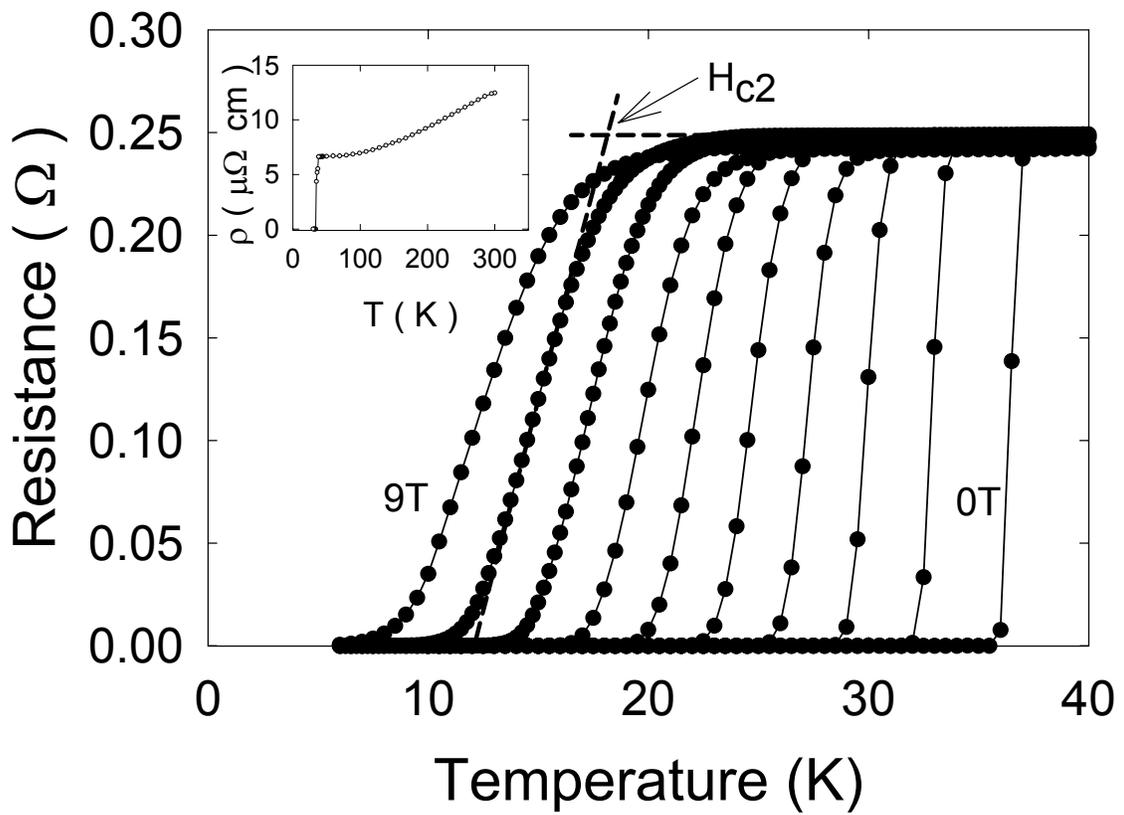

Fig. 1

Fig.2 Patnaik, S.

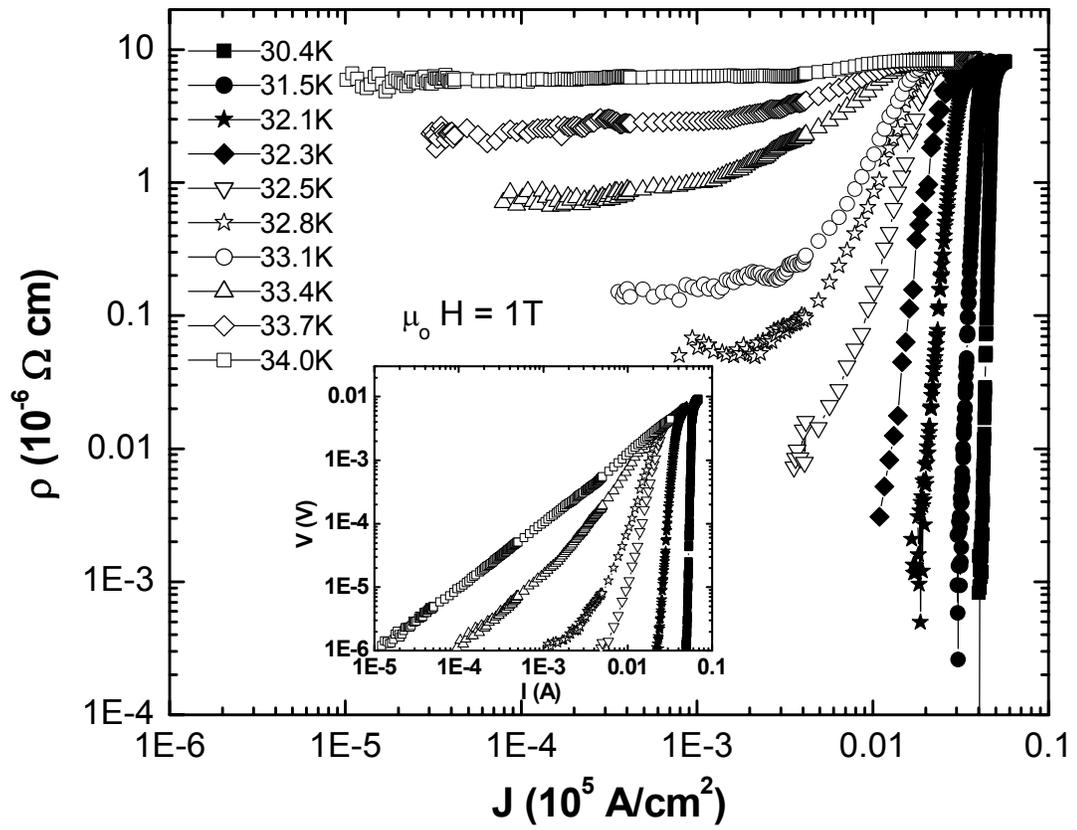

Fig. 2

Fig.3 Patnaik, S.

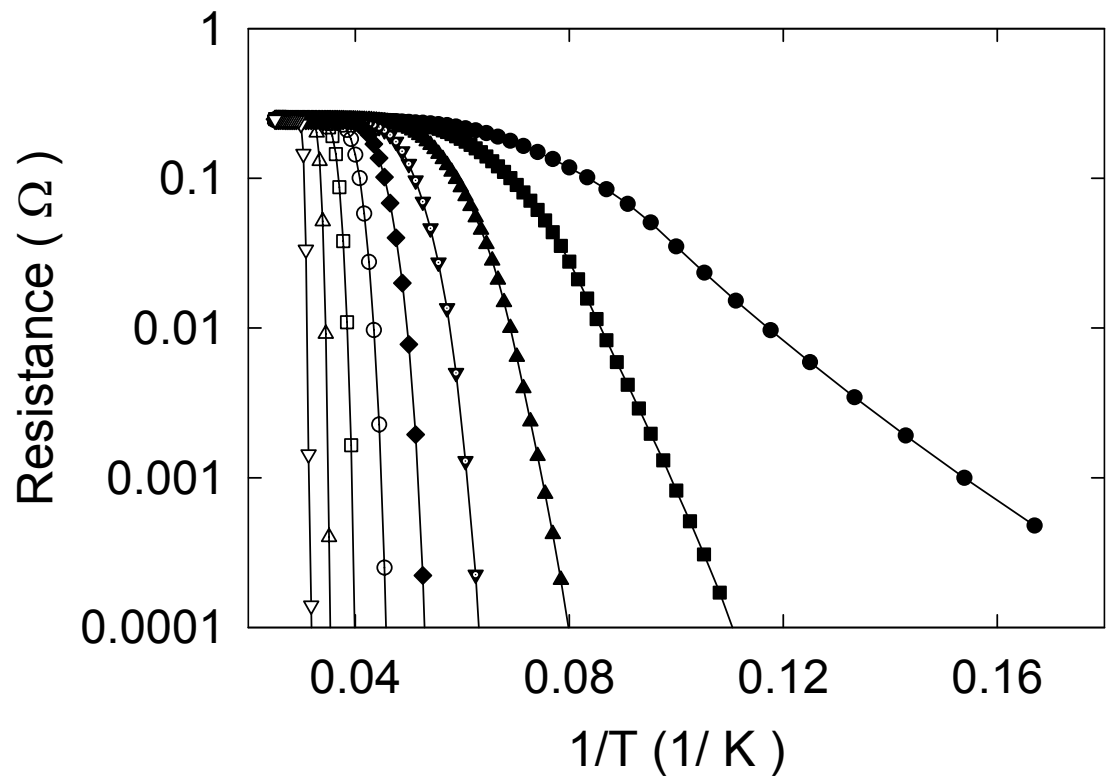

**Fig.3**

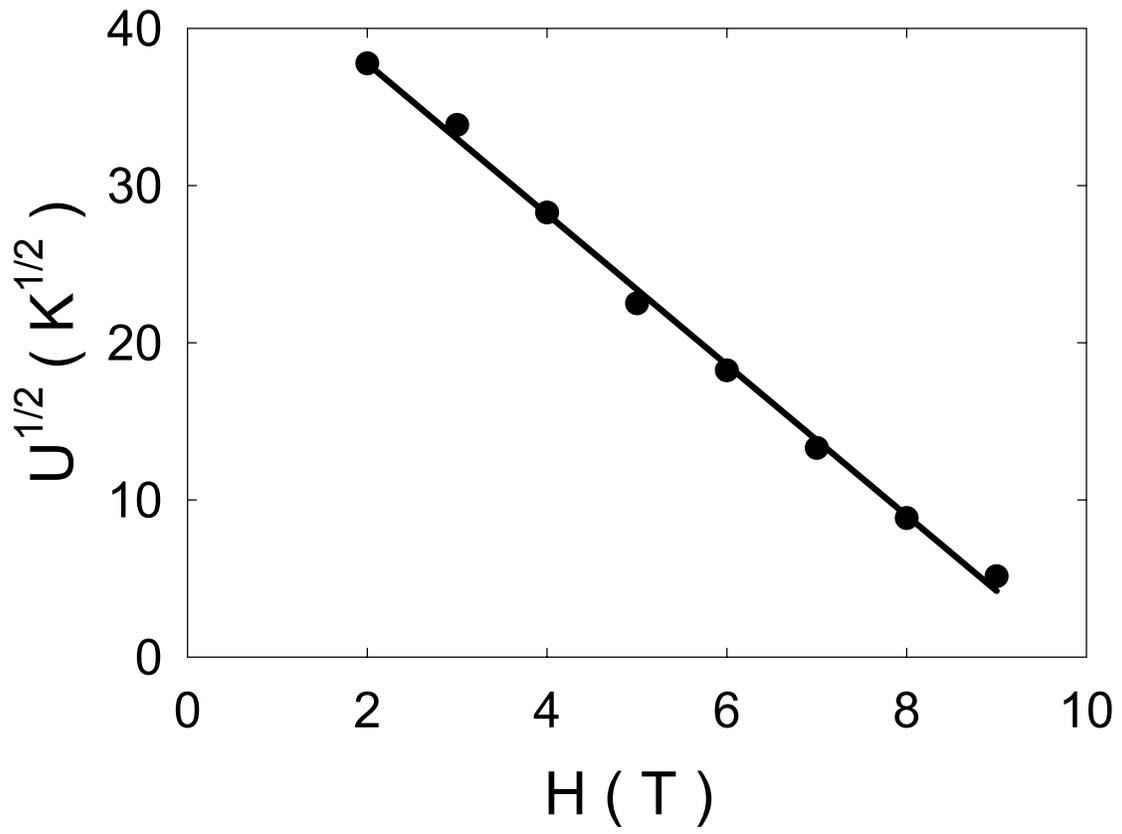

Fig.4

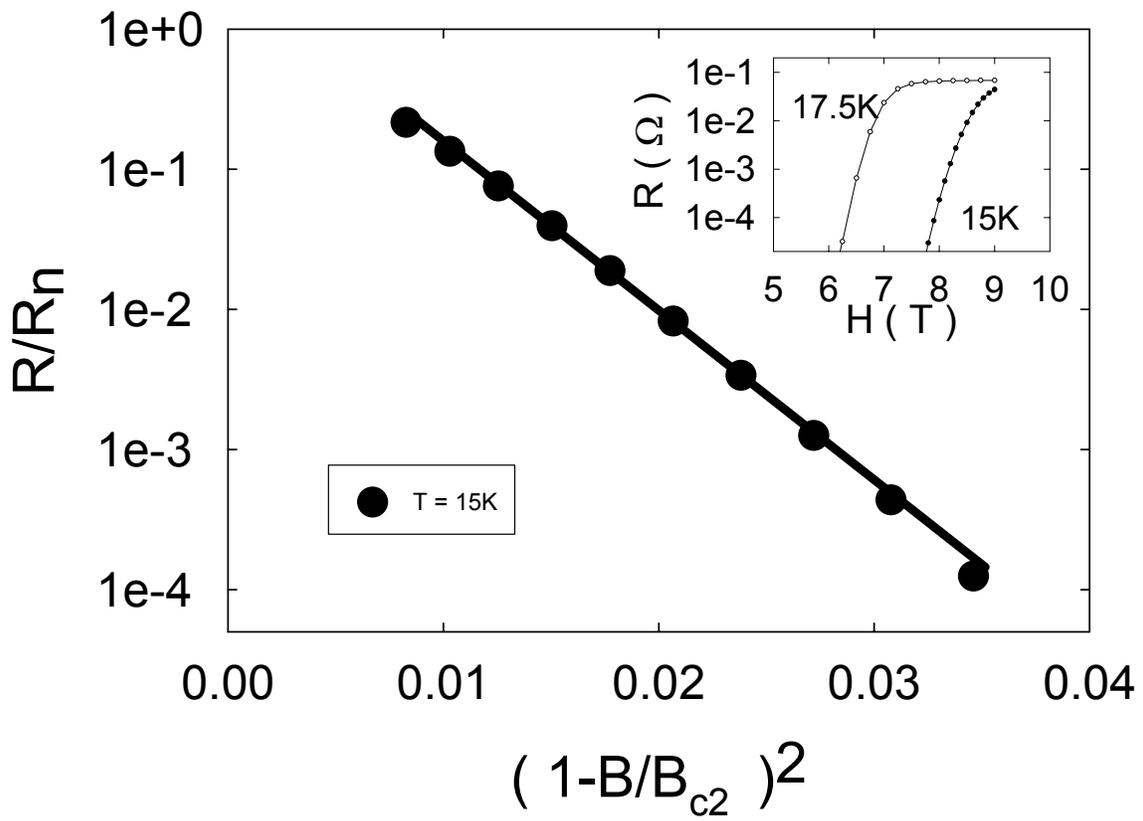

**Fig.5**

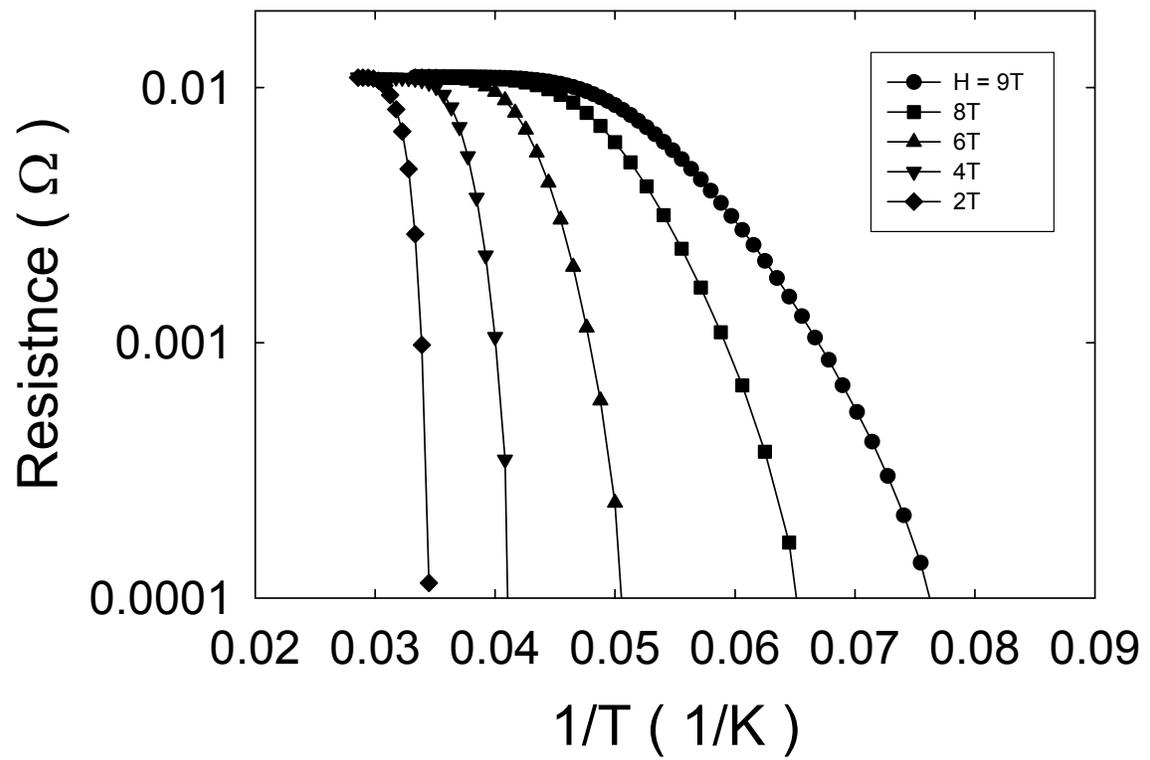

**Fig. 6**

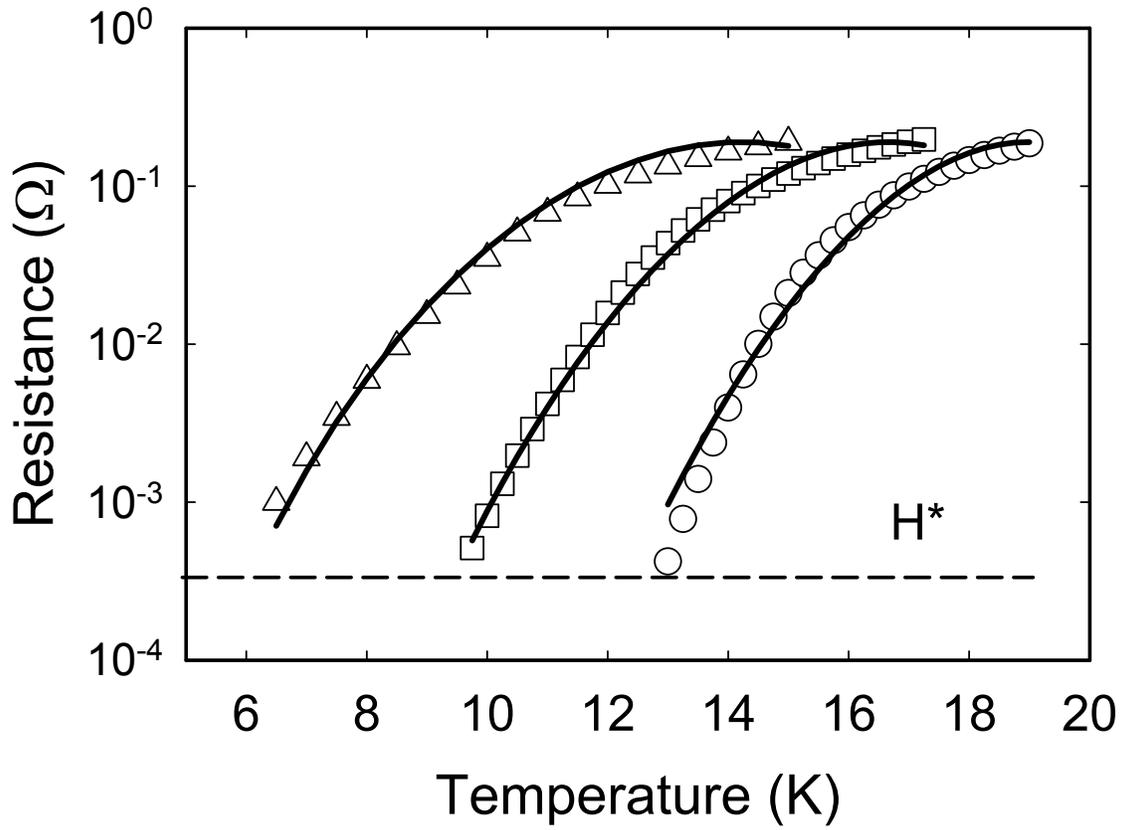

Fig. 7

Fig.8 Patnaik, S.

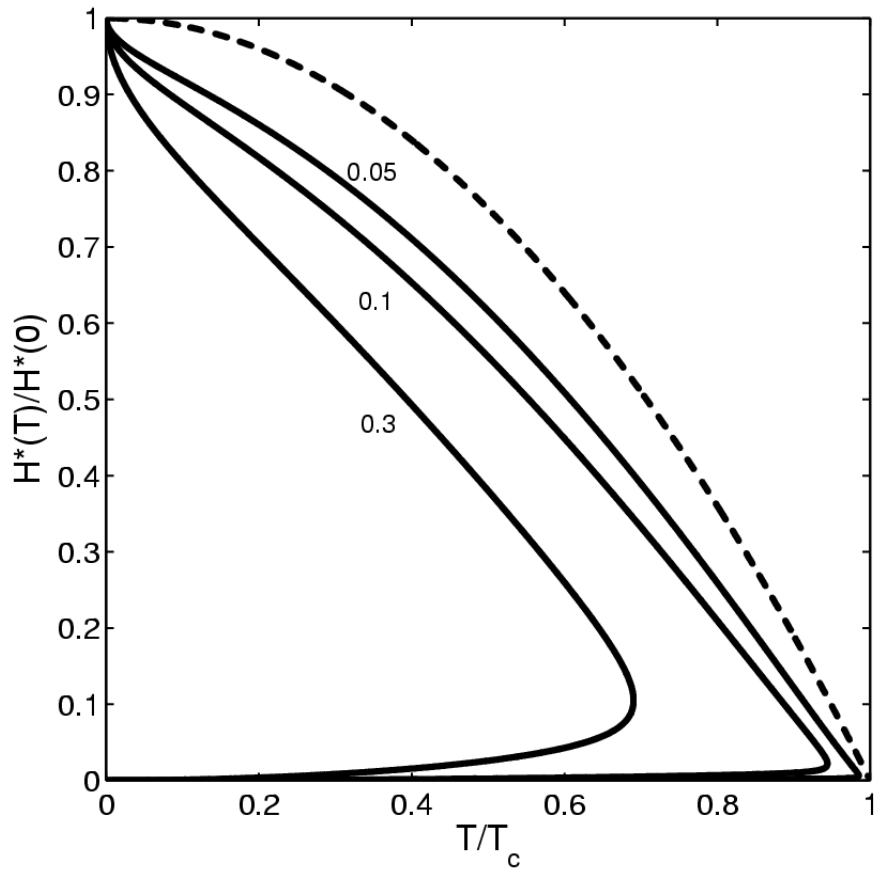

Fig. 8